\documentclass[journal]{rmaa}
\usepackage{caption}

\begin{document}

\title{CCD time-series photometry of variable stars in globular clusters and the
metallicity dependence of the horizontal branch luminosity} 

\author{
  A. Arellano Ferro,\altaffilmark{1}
D. M. Bramich, \altaffilmark{2}
Sunetra Giridhar \altaffilmark{3} }

\altaffiltext{1}{Instituto de Astronom\'ia, Universidad Nacional Aut\'onoma de
M\'exico, M\'exico. (armando@astro.unam.mx).}
\altaffiltext{2}{Qatar Environment and Energy Research Institute (QEERI), HBKU,
Qatar Foundation, Doha, Qatar. (dan.bramich@hotmail.co.uk).}
\altaffiltext{3}{Indian Institute of Astrophysics, Bangalore, India.
(giridhar@iiap.res.in).}

\shortauthor{A. Arellano Ferro, Bramich \& Giridhar}
\shorttitle{Metallicity dependence of Horizontal Branch Luminosity in Globular
Clusters}

\listofauthors{A. Arellano Ferro}
\indexauthor{A. Arellano Ferro}

\resumen{Describimos y hacemos un resumen del programa de fotometr\'ia en series
temporales CCD de c\'umulos globulares y el empleo
de la t\'ecnica de an\'alisis de im\'agenes diferenciales (DIA) en la
extracci\'on de curvas
de luz de gran precisi\'on, incluso en las regiones centrales super pobladas y para
estrellas m\'as brillantes que $V \sim$ 19 mag. Esta
estrategia nos ha
permitido descubrir aproximadamente 250 variables en una familia de 23 c\'umulos
seleccionados y
actualizar, y en algunos casos completar, el censo de
estrellas variables en cada sistema hasta cierta magnitud l\'imite. La metalicidad
y luminosidad de las variables RR Lyrae individuales se determina por la
descomposici\'on de Fourier de sus curvas de luz. En promedio esto nos conduce a la
metalicidad y la distancia del c\'umulo que las contiene. La familia de estrellas SX
Phe nos ha permitido calcular su relaci\'on P-L y determinar de manera independiente
la distancia al c\'umulo. Presentamos los valores medios de [Fe/H], $M_V$ y distancia
para un grupo seleccionado de c\'umulos globulares, obtenidos
exclusivamente a partir de la descomposici\'on de Fourier de las curvas de luz de
estrellas RR Lyrae y llevados a una escala homogenea sin precedentes. Discutimos la
dependencia de la luminsodidad de la Rama Horizontal a traves de la relaci\'on
$M_V$-[Fe/H] y encontramos que la relaci\'on debe ser tratada por separado para las
estrellas RRab y RRc.}

\abstract{We describe and summarize the findings from our CCD time-series photometry
of globular
clusters (GCs) program and the use of difference image analysis (DIA) in the
extraction of
very precise light curves even in the crowded central regions down to $V\sim 19$ mag.
We have discovered approximately 250 variable stars in a sample of 23 selected GCs and
have,
therefore, updated the census of variables in each system and in some cases we have
actually completed it down to a certain magnitude.
The absolute magnitude and [Fe/H]
for each individual RR Lyrae is
obtained via the Fourier decomposition of the light curve. An average of
these
parameters leads
to the distance and metallicity of the host GCs. We have also calibrated the P-L
relation
for SX Phe stars which enables an independent calculation of the cluster distance.
We present  the mean [Fe/H], $M_V$ and distance for a group of selected GCs
based exclusively on the RR Lyrae light curve Fourier decomposition technique and
set on a rather
unprecedented homogeneous scale. We also discuss the luminosity dependence of the
horizontal
branch (HB) via the $M_V$-[Fe/H] relation. We find that this relation should be
considered separately for the RRab and RRc stars.}

\keywords{globular clusters: metallicity and distance:stars: variables: RR Lyrae}

\maketitle

\section{Introduction}
\label{sec:intro}

Globular clusters (GCs) are the oldest stellar systems in the Galaxy and their
stellar
populations are a good representation of the most evolved and metal-deficient stars.
The population of variable stars in GCs is assorted but probably the most
important variables are the RR Lyrae (RRL) as they are numerous and have been
recognized as 
distance indicators since very early in the XX century when Shapley (1917) used them
to estimate the distance to some GCs and proved that the Sun is not at the center
of the Galaxy. It was from RRL in the Galactic bulge that Shapley (1939) and
Baade (1946) estimated the distance to the Galactic center. The most
complete compilation of variable stars in GCs is found in the 
Catalogue of Variable Stars in Galactic Globular Clusters described in detail by
Clement et
al. (2001) and which has been regularly updated for individual clusters under the
curation of Prof. Christine Clement\footnote{\texttt{
http://www.astro.utoronto.ca/$\sim$cclement/cat/listngc.html}}.

A major problem in the search and
monitoring of variables near the central regions of GCs is the overcrowding
and overlapping of star images, which has for a long time severely hampered 
our ability to complete the variable star census in GCs.
Since the introduction of CCD photometry and the approach of difference image
analysis (DIA; cf.\ Alard \& Lupton 1998), the study and census of variable stars in
GCs
has experienced a revitalization and many new variables as well as new properties
of individual cases have been discovered. Since 2002 we have been carrying out a
systematic CCD imaging program to obtain time-series photometry of a sample of GCs
with a wide
range of metallicities, with the
aim of updating the variable star census and employing the Fourier decomposition of
RRL
light curves for the determination of specific stellar physical parameters of
astrophysical
relevance, such as metallicities, luminosities, masses and radii.
The virtues of DIA as a powerful tool to discover
variable stars, or unveil small amplitude variations in Blazhko RRL stars 
in the densely populated central regions of GCs, have been
demonstrated in several recent papers by our group (e.g. Arellano Ferro et al.\
2013a, 2012; Bramich et al.\ 2011; Figuera Jaimes et al.\ 2013; 
Kains et al.\ 2013). 
We used the {\tt DanDIA}\footnote{{\tt DanDIA} is built from the DanIDL library
of IDL routines available at \texttt{http://www.danidl.co.uk}}
pipeline for the data reduction process (Bramich et al.\ 2013), which includes an 
algorithm that models the convolution kernel matching the PSF
of a pair of images of the same field as a discrete pixel array (Bramich 2008).

In the present paper we summarize the variable stars found in our program for the GCs
we have studied
and describe the fundamental approach to the calculation of the physical
parameters via the Fourier decomposition of the light curves of the RRL stars. The
mean [Fe/H], $M_V$ and distance calculated from the Fourier decomposition procedure
applied homogeneously to an extended data set are listed. These results are
essential for the subsequent
discussion on the luminosity of the Horizontal Branch and its dependence on the
metallicity, i.e. the $M_V$-[Fe/H] relation which we discuss in detail in $\S$
\ref{sec:FeHMv}.

The Fourier decomposition procedure applied homogeneously to an extended data 
set have also led to a homogeneous set of distance estimates as described 
in section $\S$ \ref{distances}.

\section{Observations and reductions}
\label{sec:ObserRed}

\subsection{Observations}

The majority of the observations for our program have been carried out using the
Johnson-Kron-Cousins $V$ and $I$ filters and have been performed with the 2.0m
Himalayan Chandra Telescope (HCT) of the Indian Astronomical Observatory
(IAO), Hanle, India. We
have also used the 2.15-m telescope of the Complejo Astron\'omico El Leoncito
(CASLEO), San~Juan, Argentina, and the Danish 1.54 m telescope
at La Silla, Chile, and the LCOGT 1 m telescopes network at the South African
Astronomical Observatory (SAAO) in Sutherland, South
Africa, at the Siding Spring Observatory (SSO) in New South Wales, Australia, and at
Cerro Tololo Inter-American Observatory (CTIO), Chile.

\subsection{Transformation to the Standard System}
\label{DIA}

Standard stars in the field of the clusters are mainly taken from the work of 
Stetson (2000)\footnote{%
 \texttt{http://www3.cadc-ccda.hia-iha.nrc-cnrc.gc.ca/\\
community/STETSON/standards}}.
Typically between 30 and 200 standard stars per GCs are used to transform our
instrumental
system into the Johnson-Kron-Cousins photometric system (Landolt 1992). The
standard
minus the
instrumental magnitude differences show a mild dependence on the colour. The
transformation equations are of the form:

\begin{eqnarray}\label{eq:transV}
V &=& v + A + B (v-i),
\end{eqnarray}
\begin{eqnarray}\label{eq:transI}
I &=& i + C + D (v-i),
\end{eqnarray}

\noindent
where $V$ and $I$ are the magnitudes in the standard system, and $v$ and $i$ are in
the instrumental system. $A, B, C$ and $D$ are the corresponding transformation
coefficients.

\section{Variable Stars in our sample of Globular Clusters}

The first step in identifying the known variable stars in a given cluster is
based in the Catalogue of Variable Stars in Galactic Globular Clusters (CVSGGC)
(Clement et al.
2001).
We then explore the light curves of all stars produced by {\tt DanDIA} in the
field of our
images for each cluster, in search of signs of
variability. Towards this, we employ different statistical approaches that have been
described in detail in the paper by Arellano Ferro et al. (2013a) and that for
brevity we do not repeat here.

\begin{figure*}
\begin{center}
\includegraphics[scale=0.35]{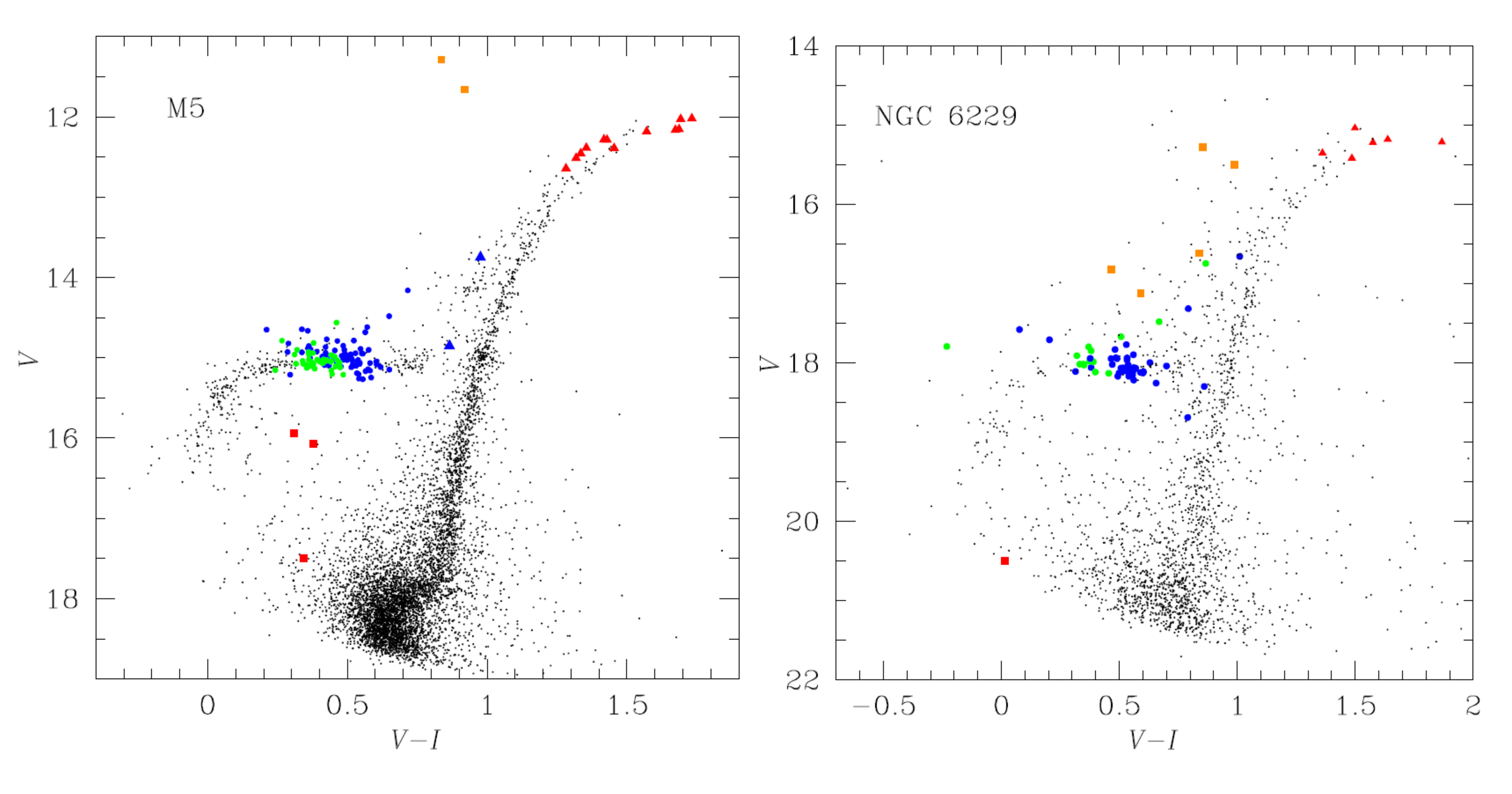}
\caption{Examples of colour-magnitude diagrams of M5 and NGC 6229 with their variable
stars marked to illustrate the regions where variable stars are expected and hence
guide the search for new variables. Colours are: blue circles- RRab, green circles-
RRc, orange squares- CW or RV Tau, red triangles- SR, red squares - SX Phe, blue
triangles- Eclipsing binaries.}
\label{CMD}
\end{center}
\end{figure*}

The colour-magnitude diagrams (CMD) of M5 and NGC~6229 shown in Fig. \ref{CMD}
illustrate 
all the regions where variable stars are expected to be present and hence the CMD
guides the search for new
variables.

\begin{table*}
\scriptsize
\begin{center}
\caption{Number of presently known variables per cluster for the most common variable
types, in a sample of GCs studied by our group$\dagger$. }
\label{NoVARS}
\begin{tabular}{lccccccccc}
\hline
GC & RRab&RRc&RRd& SX Phe&EB&CW-(AC)-RV&SR, L& Total per cluster&Ref.\\
NGC (M) & & && & & & &  &\\
\hline
288  &0/1&0/1&0/0&0/8 &0/1&0/0&0/1&0/12&1\\
1904 (M79)&0/6&1/5&0/0&0/5&0/1&0/1&0/14&1/32&2\\ 
3201 &0/72&0/7&0/0&3/24&0/11&0/0&0/8&3/122&3\\ 
4147 &0/5&1/10&0/0&0/0&0/0 &0/0&0/0&1/15&4\\   
4590 (M68)&0/14 &0/16&0/12&4/6&0/0&0/0&0/0&4/48&5\\   
5024 (M53)&0/29&2/35&0/0&13/28&0/0&0/0&1/12&16/104&6,7\\ 
5053 &0/6 &0/4&0/0&0/5&0/0&0/0 &0/0&0/15&8\\
5466 &0/13 &0/8&0/0&0/9&0/3&0/1&2/2&2/36&9\\
5904 (M5)&2/91&1/39&0/0&1/6&1/3&0/2 &11/12 &16/153&17,18 \\
6229 &14/42 &6/15 &0/0 &1/1&0/0 &2/5&6/6 &29/69&19\\
6333 (M9)&1/10 &2/10&1/1&0/0&3/4&1/1&7/8&15/34&10\\ 
6366 &0/1 &0/0&0/0&1/1 &1/1&1/1&3/4&6/8&11\\ 
6388&1/11&2/15&0/0&0/1&0/10 &3/13&40/53&46/103&21\\
6401$\ddagger$ &2/23&1/10&1/1&0/0&0/0&1/1&0/0&5/35&20 \\
6441&3/48 &0/19&1/1& 0/1&0/19&7/8&44/78&55/174&21\\
6528&1/1&1/1&0/0&0/0&1/1&0/0&4/4&7/7&21\\
6638&9/10&4/17&0/0&0/0&0/0&0/0&3/3&16/30&21\\
6652&0/3&1/2&0/0&0/0&0/1&0/0&0/2&1/8&21\\
6981 (M72)&8/37 &3/7&0/0&3/3&0/0&0/0&0/0&14/47&12\\
7078 (M15)&0/65 &0/64&0/32&0/4&0/2&0/3&0/3&0/173&13\\
7089 (M2)&5/23 &3/15&0/0&0/0&0/0&0/4&0/0&8/42&14\\
7099 (M30)&0/4&2/3&0/0&2/2&1/6&0/0&0/0&5/15&15\\
7492 &0/1&0/2&0/0&2/2&0/0&0/0&1/2&3/7&16\\
\hline
Total per type &46/516&30/305&3/47&30/106&7/63&15/40&122/212&253/1289\\
\hline
\end{tabular}
\center{\quad $\dagger$. The variable star types are adopted from the General
Catalog of Variable Stars (Kazarovets et al. 2009; Samus et al. 2009). Entries
expressed as
M/N indicate the M variables found or reclassified by our program and the total number
N of known variables. Column 12 indicates the relevant papers on the clusters by our
team.
$\ddagger$. The variable stars counting for this cluster refers to stars that
are
likely cluster members. Also 1 eclipsing binary (E1) and 12 new SR's or LPV's were
identified but are probably non-members.

\quad References: 1. Arellano Ferro et al.
(2013b); 2. Kains et al. (2012); 3.
Arellano Ferro et al. (2014a); 4.
Arellano Ferro et al. (2004); 5. Kains et al. (2015), 6. Arellano Ferro et al.
(2011), 7. Bramich et al. (2012), 8. Arellano Ferro et al. (2010), 9. Arellano Ferro
et al. (2008a), 10. Arellano
Ferro et al.
(2013a), 11. Arellano Ferro et al. (2008b), 12. Bramich et al. (2011); 13.
Arellano Ferro et al. (2006); 14. L\'azaro et al. (2006); 15. Kains et al. (2013); 
16. Figuera Jaimes et al. (2013); 17. Arellano Ferro et al. (2015a), 18. Arellano
Ferro et al. (2016); 19. Arellano Ferro
et al. (2015b); 20. Tsapras et al. (2017); 21. Skottfelt et al.
(2015).}
\end{center}

\end{table*}

We have so far discovered 253 new variables of different types
(46 RRab, 30 RRc and 3 double mode RRL stars, 30 SX Phe, 7 eclipsing binaries (EB), 15
Pop II 
Cepheids (of which 2 are anomalous) and 122 SR or L as listed in
Table \ref{NoVARS}.

\section{Physical parameters of RR~Lyrae stars}
\label{sec:Four}

The Fourier decomposition of the RRL light curves is performed by fitting the
observed light curve in $V$ with a Fourier series model of the form:

\begin{equation}
\label{eq.Foufit}
m(t) = A_0 + \sum_{k=1}^{N}{A_k \cos\ ({2\pi \over P}~k~(t-E) + \phi_k) },
\end{equation}

\noindent
where $m(t)$ is the magnitude at time $t$, $P$ is the period, and $E$ is the epoch. A
linear
minimization routine is used to derive the best-fit values of the 
amplitudes $A_k$ and phases $\phi_k$ of the sinusoidal components. 
From the amplitudes and phases of the harmonics in eq.~\ref{eq.Foufit}, the 
Fourier parameters, defined as $\phi_{ij} = j\phi_{i} - i\phi_{j}$, and $R_{ij} =
A_{i}/A_{j}$, are computed. 

Subsequently, the low-order Fourier parameters can be used in combination with
semi-empirical calibrations to calculate [Fe/H] and $M_V$ for
each RRL and hence the mean values of the metallicity and distance for the host
cluster.

\subsection{[Fe/H] and $M_V$ calibrations}
\label{calibrations}
For the calculation of [Fe/H] we adopted the following calibrations:

\begin{eqnarray}
\label{eq:RR0_Fe}
{\rm [Fe/H]}_J= - 5.038 - 5.394 P + 1.345 \phi^{(s)}_{31},
\end{eqnarray}

\begin{eqnarray}
\label{eq:RR1_Fe}
{\rm [Fe/H]}_{ZW} = 52.466 P^2 - 30.075 P + 0.131 \phi^{(c)2}_{31}  \nonumber \\
 - 0.982 \phi^{(c)}_{31} - 4.198 \phi^{(c)}_{31} P + 2.424
\end{eqnarray}

\noindent
from Jurcsik \& Kov\'acs (1996) and Morgan et al. (2007) for RRab and RRc stars,
respectively. The iron abundance on the Jurcsik \& Kov\'acs (1996) scale can be
converted into the Zinn \& West (1984) scale using the equation
[Fe/H]$_J$ = 1.431[Fe/H]$_{ZW}$ + 0.88 (Jurcsik 1995).

For the calculation of $M_V$ we adopted the following calibrations:

\begin{equation}
\label{eq.RR0_Mv}
M_V= -1.876~log P -1.158 A_1 + 0.821 A_3 + 0.41
\end{equation}

\begin{equation}
\label{eq.RR1_Mv}
M_V= -0.961 P - 0.044 \phi^{(s)}_{21} -4.447 A_4 + 1.061
\end{equation}

\noindent
from Kov\'acs \& Walker (2001) and Kov\'acs (1998) for the RRab and RRc stars,
respectively. The zero points of eqs. \ref {eq.RR0_Mv} and \ref {eq.RR1_Mv} have been
calculated to scale the
luminosities of RRab and RRc stars to the distance modulus of 18.5 mag for the Large
Magellanic Cloud (LMC) (see the discussion in $\S$ 4.2 of Arellano Ferro et al.
2010).

In the equations above, the superscript ($c$) indicates that the Fourier
decomposition is
done using a cosine series, as in eq. \ref{eq.Foufit}, whereas a superscript ($s$)
means that the
equivalent sine series was employed.

In Table \ref{MV_FEH:tab} we list the GCs studied by our team and the resulting
[Fe/H]$_{ZW}$ and
$M_V$ estimated via the Fourier decomposition of the light curves of the RRab and RRc
stars.
We have also included the two metal-rich clusters NGC 6388 and NGC 6441 studied by
Pritzl et al. (2001; 2002), NGC 1851 (Walker 1998) and NGC 5272 (Cacciari et al.
2005) since the light curve Fourier decomposition parameters are available in those
papers and hence we can include them in our homogeneous calculation of their physical
parameters via the equations \ref{eq:RR0_Fe} through \ref{eq.RR1_Mv}. The use of the
above equations and their zero points form the basis of the discussion of the
$M_V$-[Fe/H]
relation and the cluster distances on a homogeneous scale, which we present in later
sections for the clusters in our sample.

\section{The $M_V$-[{F\MakeLowercase{e}/H}] relation}
\label{sec:FeHMv}

The importance of RR Lyrae stars of being good distance indicators is well known since
the early XXth century. Shapley
(1917) recognized that "The median magnitude of the
short-period variables [RR Lyrae stars] apparently has a rigorously constant value in
each globular cluster", a fact that was used later by Shapley himself to describe the
Galactic distribution of globular clusters (Shapley 1918).
This apparently constant value of the mean magnitude of the RR Lyraes can now be
interpreted
as the luminosity level of the horizontal branch (HB) being constant in all
globular clusters. The fact that this is not exactly the case, but instead that
metallicity plays a role in the luminosity level of the HB, 
has been demonstrated via synthetic models of 
  Lee, Demarque \& Zinn (1990) although the metallicity dependence was
 also known through the work by Sandage(1981a,b). Lee, Demarque \& Zinn (1990)
  provide a calibration of the $M_V$-[Fe/H] relation and discuss its dependence on 
  helium abundance. This work was followed by more empirical calibrations
  by Walker(1992), Carney et al. (1992) and Sandage(1993).
Complete summaries on the
calibration of the $M_V$-[Fe/H] relation can be found in the works
of Chaboyer (1999), Cacciari \& Clementini (2003) and Sandage \& Tammann (2006). 
While the relation is believed to be linear in empirical work, a non-linear
nature is advocated by theoretical work, e.g.  Cassisi et al. (1999)
and VandenBerg et al. (2000).

\begin{table*}
\scriptsize
\begin{center}
\caption{Mean values of [F\MakeLowercase{e}/H] and $M_V$ from a homogeneous
Fourier decomposition of
the light curves of RR Lyrae cluster members.$^1$}
\label{MV_FEH:tab}
\begin{tabular}{lc|ccc|ccc|c}

\hline
GC & Oo Type&[Fe/H]$_{ZW}$ & $M_V$&N&[Fe/H]$_{ZW}$ & $M_V$&N&Ref.\\
\hline
NGC (M) & &RRab& & & RRc& & &\\
\hline
1851      &I &-1.437$\pm$0.098&0.540$\pm$0.026 &10&-1.397$\pm$0.130&0.586$\pm$0.019& 5
&23\\
3201      &I &-1.483$\pm$0.098&0.604$\pm$0.045 & 19&-1.473$\pm$0.098&0.576$\pm$0.045 &
2 &3 \\
4147      &I &-1.516$\pm$0.038&0.583$\pm$0.066 & 2 & -1.675$\pm$0.260&0.562$\pm$0.064&
6 &  4\\
5272 (M3)  &I &-1.560$\pm$0.156&0.589$\pm$0.046 &59
&-1.648$\pm$0.136&0.555$\pm$0.059&23 &24\\
5904 (M5)  &I &-1.444$\pm$0.094&0.577$\pm$0.081  &
35&-1.490$\pm$0.106&0.575$\pm$0.028&22 &19\\
6171 (M107)&I &-1.310$\pm$0.120&0.635$\pm$0.104  & 6 &-1.034$\pm$0.115&0.573$\pm$0.037
&7 &22\\
6229      &I &-1.416$\pm$0.065&0.621$\pm$0.050  & 12&-1.401$\pm$0.160&0.564$\pm$0.078
&7 &20 \\
6362      &I
&-1.345$\pm$0.149&0.589$\pm$0.032&6&-1.229$\pm$0.221&0.574$\pm$0.053&10&this work \\
6366      &I &-0.844&0.705  & 1 &-- &-- &-- &11$^2$ \\
6401      &I &-1.254$\pm$0.064&0.648$\pm$0.062  &22 &-1.266$\pm$0.225
&0.575$\pm$0.034&9& 21\\
6934   &I &-1.528$\pm$0.126 &0.567$\pm$0.067&10 &-1.533$\pm$0.184
&0.591$\pm$0.086 &8
&this work\\
6981 (M72) &I &-1.482$\pm$0.030&0.623$\pm$0.023  &13 &-1.661$\pm$0.082&0.568$\pm$0.038
& 4&14\\
\hline
NGC(M)  & &RRab& & & RRc& & &\\
\hline
288       & II&-1.852$^5$&  0.376& 1 &-1.591  &0.579& 1 &1\\
1904 (M79) & II&-1.854$\pm$0.137$^5$&  0.459$\pm$0.084& 5 &-1.732 &0.584 & 1 &2
\\
4590 (M68) & II&-2.085$\pm$0.093$^5$&  0.495$\pm$0.067& 3 &-2.087$\pm$0.026
&0.532$\pm$0.011&15 &5 \\
5024 (M53) & II&-1.956$\pm$0.066$^5$&  0.452$\pm$0.052 &19
&-1.839$\pm$0.131&0.519$\pm$0.062& 3 &6 \\
5053      & II&-2.085$\pm$0.159$^5$&  0.462$\pm$0.081 & 3
&-1.995$\pm$0.184&0.550$\pm$0.048& 4 &7 \\
5466      & II&-2.051$\pm$0.139$^5$&  0.438$\pm$0.088 & 7 &-1.900$\pm$0.212 
&0.532$\pm$0.065 & 5 &8 \\
6333 (M9)  & II&-1.912$\pm$0.130$^5$&  0.467$\pm$0.034& 7 &-1.709$\pm$0.230 
&0.554$\pm$0.042 & 6 & 9\\
6341 (M92) & II&-2.125$\pm$0.120$^5$&   0.498$\pm$0.066& 5 &-2.063$\pm$0.019
&0.504$\pm$0.017& 2 &10 \\
7078 (M15) & II&-2.240$\pm$0.188$^5$&  0.506$\pm$0.044 &14 &-2.097$\pm$0.070
&0.524$\pm$0.031 & 8 &15 \\
7089 (M2)  & II&-1.775$\pm$0.179$^5$&  0.552$\pm$0.150& 11 &-1.760$\pm$0.165
&0.507$\pm$0.067& 2 &16 \\
7099 (M30) & II&-2.066$\pm$0.050$^5$&  0.405$\pm$0.044& 3 &-2.032  &0.541& 1 &17
\\
7492      & II&-1.89$^{3,5}$&  0.376 & 1 &--&--&--&18$^4$ \\
\hline
 & &RRab& & & RRc&  & &\\
\hline
6388  &III&-1.345$\pm$0.054&0.528$\pm$0.040&2 &-0.672$\pm$0.236&0.609$\pm$0.068&6
&12 \\
6441  &III&-1.348$\pm$0.172&0.434$\pm$0.078&7 &-1.028$\pm$0.338&0.546$\pm$0.080&8
&13 \\

\hline

\end{tabular}
\center{\quad Notes: $^1$ Quoted uncertainties are 1-$\sigma$
errors calculated from the scatter in the data for each cluster. The number of stars
considered in the calculations is given by N. $^2$. The
only RRL V1 is probably not a cluster member. $^3$.
Adopted since published Fourier coefficients are insufficient.  $^4$ Based on
one light curve not fully covered. $^5$ This value has a -0.21 dex added, see
$\S$ \ref{sec:FeHMv} for a discussion.\\

\quad References are the source of the Fourier coefficients: 1. Arellano Ferro et al.
(2013b); 2. Kains et al. (2012); 3.
Arellano Ferro et al. (2014a); 4.
Arellano Ferro et al. (2004); 5. Kains et al. (2015), 6. Arellano Ferro et al.
(2011), 7. Arellano Ferro et al. (2010), 8. Arellano Ferro et al. (2008a), 9. Arellano
Ferro et al.
(2013a), 10. Mar\'in (2002), 11. Arellano Ferro et al. (2008b), 12. Pritzl, et al.
(2002), 13. Pritzl, et al. (2001), 14. Bramich et al. (2011); 15.
Arellano Ferro et al. (2006); 16. L\'azaro et al. (2006); 17. Kains et al. (2013); 18.
Figuera Jaimes et al. (2013); 19. Arellano Ferro et al. (2016); 20. Arellano Ferro et
al. (2015b); 21. Tsapras et al. (2017); 22. Clement \& Shelton(1997); 23. Walker
(1998); 24. Cacciari et al. (2005).}
\end{center}

\end{table*}

\begin{figure*}
\begin{center}
\includegraphics[scale=1.1]{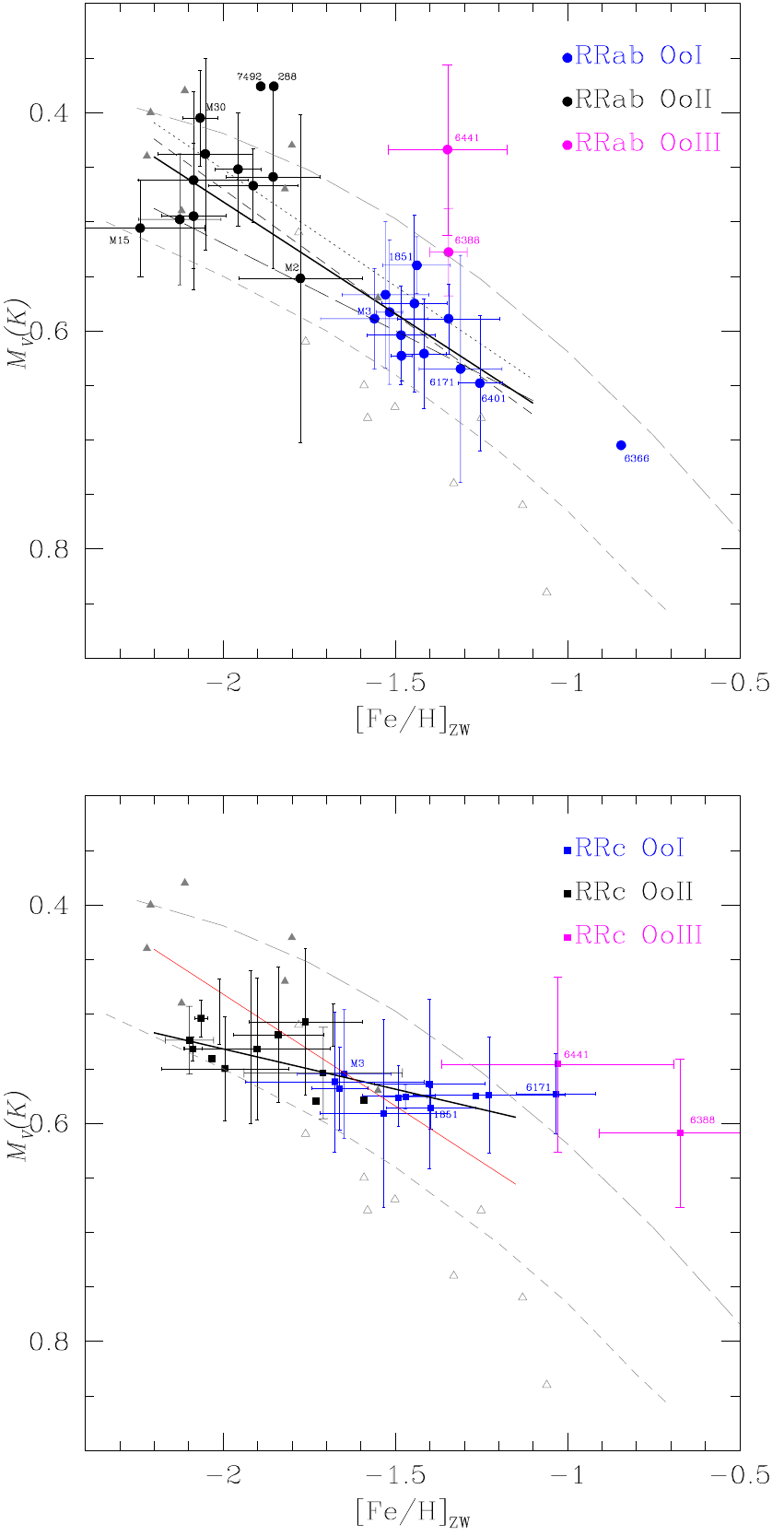}
\caption{Empirical Mv-[Fe/H] relation from the homogeneous Fourier decomposition of RR
Lyrae light curves in families of OoI, OoII and OoIII globular clusters. Data points
are plotted with 1-$\sigma$ error bars. The upper panel shows the cluster
distribution 
where $M_V$ and [Fe/H] were calculated from the light curve decomposition of RRab
stars. The solid black straight line is the least-squares fit to the data (except the
OoIII clusters, and NGC~288, NGC~6366 and NGC~7492 which are based only on one star)
and
corresponds to eq. \ref{eq:MvFeH_0}. Dotted, short, and long dashed straight lines are
the
relations of Clementini et al. (2003), Chaboyer (1999) and that adopted by Harris
(1996) respectively.
The lower panel is for the case of RRc stars and corresponds to eq. \ref{eq:MvFeH_1}. 
The red straight line is the reproduction 
of the calibration for the RRab stars from the upper panel to stress the
significantly 
different slope. Two non-linear theoretical calibrations from Cassisi et al. (1999)
(long dash) and VandenBerg et al. (2000) (short-dash) are shown. 
The cluster distribution of Caputo et al. (2000) (gray triangles) is also included
for reference.}
\label{MvFE}
\end{center}
\end{figure*}

The metallicity dependence of the Horizontal Branch (HB) luminosity can be studied
from the mean
absolute
magnitude $M_V$ and [Fe/H] obtained for individual globular clusters using the Fourier
decomposition of RRL light curves 
and the semi-empirical calibrations described in $\S$ \ref{calibrations}. Numerous
Fourier decompositions of RRL light curves can be 
found in the literature. However, over the years, each author has used different
calibrations and zero points to estimate $M_V$ and [Fe/H]. Our own group has
also
used slightly different equations in the earlier papers and it was not until
the work by Arellano Ferro et al. (2010) that the firm zero points in eqs.
\ref{eq.RR0_Mv} and \ref{eq.RR1_Mv} were adopted and we have used them subsequently. 

To achieve a homogeneous approach, we have taken the light curve 
Fourier decomposition coefficients given 
in the original papers wherever they are available, (i.e. always in the case for
our own works, 
and also in a few papers by other authors as indicated in Table \ref{MV_FEH:tab}). 
We have then carried out the calculations using the equations in $\S$
\ref{calibrations} for as
many RRL's 
as available in each cluster. In Table \ref{MV_FEH:tab}, the final mean values of
$M_V$ and [Fe/H] and the adopted reddening for each case are recorded.
The table
is organized by Oosterhoff types; Oosterhoff (1939, 1944) realized that the periods of
RRab stars in a given cluster group around two values;
0.55d (Oosteroff type I or OoI) and 0.65d (Oosteroff type II or OoII). OoI clusters
are systematically more metal rich than OoII clusters. A third Oosterhoff class 
(OoIII) (Pritzl et al. 2000), which presently contains only two GCs, NGC 6388 and NGC
6441, is represented by very metal-rich systems where the periods of their RRab
stars average about 0.75d. In Table \ref{MV_FEH:tab} we include
of 12 OoI, 12 OoII and 2 OoIII clusters. The calculations have been
performed independently for RRab and RRc stars. For clusters with differential
reddening, i.e. NGC 3201, NGC 6333 and NGC 6401, care has been taken
in calculating the individual reddenning for each RRL. The interested reader is
referred to the original papers for a detailed discussion on that subject. 

It is known that eq. \ref{eq:RR0_Fe} overestimates [Fe/H] for metal poor clusters.
This problem has been addressed by Jurcsik \& Kov\'acs (1996), 
Schwarzenberg-Czerny \& Kaluzny (1998), Kov\'acs (2002), Nemec (2004) and Arellano
Ferro et al. (2010).
It is difficult to quantify a correction to be applied, and this is likely 
a function of the metallicity, however, empirical estimations in the above papers
point to a value between --0.2 and  --0.3 dex on the scale of eq. \ref{eq:RR0_Fe}. We
have adopted --0.3 dex, which on the $ZW$ scale is equivalent to --0.21 dex.
Therefore,
the values listed in Table \ref{MV_FEH:tab} for clusters with [Fe/H]$_{ZW} <$ --1.5
dex
were obtained by adding --0.21 dex to the value of [Fe/H]$_{ZW}$ found via eq.
\ref{eq:RR0_Fe}.

In Fig. \ref{MvFE} we show the distribution of clusters in the $M_V$-[Fe/H]
plane obtained from the RRab stars
(left panel) and the RRc stars (right panel). In the Figure we have included as
reference, in gray colour, two theoretical versions of the $M_V$-[Fe/H] relation of
Cassisi 
et al. (1999) and VandenBerg et al. (2000) and the semi-empirical cluster distribution
of Caputo et al. (2000) with their [Fe/H] values converted into the ZW scale.

The solid black lines are the linear fits to the data and correspond to the equations:

\begin{eqnarray}\label{eq:RR0_Mv}
\label{eq:MvFeH_0}
M_V= 0.205 (\pm 0.025) {\rm [Fe/H]}_{ZW}  + 0.892 (\pm 0.045),
\end{eqnarray}

\noindent
for the RRab solutions, and

\begin{eqnarray}\label{eq:RR1_Mv}
\label{eq:MvFeH_1}
M_V= 0.074 (\pm 0.015) {\rm [Fe/H]}_{ZW}  + 0.680 (\pm 0.025),
\end{eqnarray}

\noindent
for the RRc solutions.

In the above fits we have omitted the clusters NGC 6388 and NGC 6441 which have been
classified as of the Oo III type (Catelan 2009), also NGC 288 and NGC 7492, 
where the analysis is based only on one star, and NGC 6366 since it has been
argued by Arellano
Ferro et al. (2008b) that the sole reported RRab star is not a cluster member.

Eq. \ref{eq:MvFeH_0} can be compared with the other well known calibrations, e.g.
$M_V=0.23 (\pm 0.04) {\rm [Fe/H]} + 0.93 (\pm 0.12)$ of Chaboyer (1998) (short
dashed line in Fig. \ref{MvFE}), $M_V=0.22 (\pm 0.05) {\rm [Fe/H]} + 0.89 (\pm 0.07)$
of Graton et al. (2003) or $M_V=0.214 (\pm 0.047) {\rm [Fe/H]} + 0.88
(\pm
0.07)$ of Clementini et al. (2003) (dotted line), to which
our relation is consistent within the respective
uncertainties. It may also be compared with the relation $M_V=0.16 {\rm [Fe/H]} +
0.84$ adopted by Harris
(1996) (long dashed line in Fig. \ref{MvFE})or the one obtained by Kains et al. (2012)
$M_V = 0.16 (\pm 0.01) {\rm [Fe/H]}_{ZW} + 0.85 (\pm 0.02)$
which are considerably flatter. We may comment at this point that the relation of
Kains et al. (2012) was calculated taking $M_V$ and [Fe/H]$_{ZW}$ from the
literature and that, although an attempt was made to set $M_V$ and [Fe/H]$_{ZW}$ into
homogeneous scales, those parameters may have been calculated from different versions
of the Fourier parameters calibrations and zero points. Also this relation was
calculated
including both RRab and RRc stars.

The most striking feature of Fig. \ref{MvFE} is that the cluster distribution for
the RRc stars (lower panel)
is significantly flatter than the distribution for the RRab stars,
and that it has a much smaller scatter. Eqs. \ref{eq:RR0_Mv} and \ref{eq:RR1_Mv}
intersect at [Fe/H]$_{ZW}=-1.62$ but the differences at [Fe/H]$_{ZW}=-2.2$ and
[Fe/H]$_{ZW}=-1.1$ are as large as 0.076 dex and 0.068 dex respectively.
We note that a visual inspection of Kains et al. (2012) Figure 9 may also suggest
different slopes for the RRab and RRc stars.

We conclude that the two distributions are authentically different, and that RRab and
RRc stars should be treated separately when considering the $M_V$-[Fe/H] relation.

\begin{table*}
\scriptsize
\begin{center}

\caption{Distances for a sample of Globular Clusters estimated homogeneously
from the RR Lyrae Fourier decompositions. Distances for the SX Phe are
calculated from their P-L relation.}
\label{distance}

\begin{tabular}{lccccccc}
\hline
GC &$d (kpc)$&$d (kpc)$& $d$ (kpc) &No. of & $d$ (kpc)&$E(B-V)$ &Ref. \\
NGC(M)  & (RRab)& (RRc) & (SX Phe) &SX Phe&  (SX Phe)& &\\
   &  &   &P-L AF11 &&P-L CS12 &&\\
\hline
 288  &9.0$\pm$0.2 &-- &8.8$\pm$0.4&6&9.4$\pm$0.6&0.03 & 1 \\
1851 &12.6$\pm$0.2 &12.4$\pm$0.2&--&--& --&0.02&this work\\
1904 (M79) &13.3$\pm$0.4&12.9&--& --&--&0.01& 2\\ 
3201 &5.0$\pm$0.2&5.0$\pm$0.1&4.9$\pm$0.3&16&5.2$\pm$0.4&dif& 3 \\ 
4147 &19.3 &18.7$\pm$0.5 &--& --&--&0.02&4\\   
4590 (M68)&9.9$\pm$0.3&10.0$\pm$0.2&9.8$\pm$0.5&6 &--&0.05& 5\\
5024 (M53)&18.7$\pm$0.4&18.0$\pm$0.5&18.7$\pm$0.6&13&20.0$\pm$0.8&0.02&6 \\ 
5053 &17.0$\pm$0.4 &16.7$\pm$0.4&17.1$\pm$1.1&12&17.7$\pm$1.2&0.02&7 \\
5466 &16.6$\pm$0.2 &16.0$\pm$0.6&15.4$\pm$1.3&5&16.4$\pm$1.3&0.00&8 \\
5904 (M5) &7.6$\pm$0.2 &7.5$\pm$0.3&6.7$\pm$0.5&3&7.5$\pm$0.2&0.03&19 \\
6229 &30.0$\pm$1.5  &30.0$\pm$1.1 &27.9 & 1&28.9&0.01&20 \\
6333 (M9) &8.1$\pm$0.2 &7.9$\pm$0.3&--&--& --&dif&9\\
6341 (M92)&8.2$\pm$0.2 &8.2$\pm$0.4&--&--& --&0.02&10\\
6362 &7.8$\pm$0.1&7.7$\pm$0.22&7.1$\pm$0.2&6&7.6$\pm$0.2&0.09&this work\\
6366 &3.3 &--&--&--& --&0.80&11\\
6388 &9.5$\pm$1.2&11.1$\pm$1.1 &--&--&--&0.40&this work\\ 
6401 &6.35 &-- &--&--&--&dif& 21\\
6441 &11.0$\pm$1.8 &11.7$\pm$1.0 &--&--& --&0.51&this work\\ 
6934 &15.9$\pm$0.4&16.0$\pm$0.6&15.8& 1&18.0&0.10&this work \\
6981 (M72)&16.7$\pm$0.4&16.7$\pm$0.4&16.8$\pm$1.6&3&18.0$\pm$1.0&0.06&14 \\

7078 (M15)&9.4$\pm$0.4&9.3$\pm$0.6&--& --&--&0.08& 15\\
7089 (M2) &11.1$\pm$0.6&11.7$\pm$0.02&--& --&--&0.06& 16\\
7099 (M30)&8.32$\pm$0.3 &8.1&8.0&1&8.3&0.03&17 \\
7492 &24.3$\pm$0.5&--&22.1$\pm$3.2&2& 24.1$\pm$3.7&0.00&18\\

\hline
\end{tabular}
\raggedright
\center{Notes: The distance from
the P-L relationships of SX Phe and the number of stars included are given, when
available, in columns 4, 5 and 6.\\

\quad References: As in Table 2}
\end{center}
\end{table*}

\subsection{Distances to the globular clusters}
\label{distances}

Given the mean values of $M_V$ in Table \ref{MV_FEH:tab}, we have calculated
the corresponding distances. Since $M_V$ for the RRab and the RRc come from
independent calibrations, we have calculated and reported these two values of the
distance as they are truly independent. The values are reported in Table
\ref{distance}. In column 7 we list the adopted values of $E(B-V)$ in the
calculations.
For the clusters with differential reddening the reader is referred to the original
papers for detailed discussions on the individual reddening estimations.

In the cases of clusters with SX Phe stars, we have employed the $P-L$ relation
calculated by Arellano Ferro et al. (2011) (AF11) which is of the form:

\begin{equation}
\label{SX-PLours}
M_V= -2.916~log P - 0.898.
\end{equation}

The corresponding distances and number of stars used in the calculation are listed
in columns 4 and 5 of Table \ref{distance} respectively. Note that they are in
excellent agreement with the distances from the RRab and RRc stars. The P-L
calibration of Cohen
and Sarajedini (2012) (CS12); $M_V= -3.389~logP - 1.640$ was also considered and
the results are listed in column 6 of Table \ref{distance}. This calibration tends to
produce 
distances between 5 and 15 \% larger than eq. \ref{SX-PLours}, although both
estimates agree well within the 1-$\sigma$ errors.

In our opinion the distances reported in Table \ref{distance} represent the best
set of homogeneous results that have been obtained from photometric methods for the 
sample of globular clusters studied.

\section{Conclusions}

The CCD time-series photometric study of globular clusters, in combination with
difference
image analysis, has proven to be very fruitful in the discovery of new
variables and hence in updating the variable star census in each of the studied
clusters. In many cases, more accurate periods and new ephemerides have been provided 
for some known variables.
Even in those cases where no new variables have been found, we have generally been
able to
establish magnitude limits for the existence of hitherto unknown variables.

The adopted semi-empirical calibrations and their zero points for the
calculation of physical parameters via the Fourier decomposition of RRL light
curves leads to a set of homogeneous mean values of $M_V$ and [Fe/H] for a family of
GCs with a wide range of metallicities, hence including a good sample of both 
OoI and OoII types.
These results enable a discussion of the variation of the
HB luminosity with the metallicity through the well-known
$M_V$-[Fe/H] relation, based on a rather unprecedented homogeneous approach.
The resulting $M_V$-[Fe/H] relation from the Fourier decomposition of RRab stars in a
family
of 
12 clusters reproduces the mean relation found by Chaboyer (1999) from a number of 
independent methods. However, the $M_V$-[Fe/H] relation found from the Fourier
decomposition of RRc stars
has a significantly lower slope and smaller dispersion. 
Finally we note that the metal-rich OoIII clusters do not follow the above relations.

\vskip 1.0cm
\noindent

A.A.F. acknowledges grant IN106615-17 from DGAPA-UNAM, M\'exico and D.M.B.
acknowledges
NPRP grant X-019-1-006 from the Qatar National Research Council (a member of the Qatar
foundation). We are indebted to Dr. Javier Ahumada for his valuable comments and
suggestions and to an anonymous referee for her/his pertinent corrections.

\end{document}